\documentclass{optica-article}

\journal{opticajournal} 

\articletype{Research Article}

\usepackage[version=4]{mhchem}
\usepackage{siunitx}
\usepackage{bm}
\renewcommand*{\vec}[1]{\bm{#1}}
\newcommand{\dif}{\,\mathrm d}
\newcommand\mi{\mathrm{i}}

\newcommand{\modify}[1]{{#1}}

\DeclareMathOperator{\real}{Re}
\usepackage{xcolor}

\begin{document}

\title{Size Scaling Law for Radiation Losses of Modes in Photonic Crystal Surface Emitting Devices}

\author{
Qingyuan Zhang,\authormark{1,2,3,$\dag$}
Ming L\"u,\authormark{1,$\dag$,*}
Jiang Hu,\authormark{1,4}
Qi Dai,\authormark{1}
Rui Zhu,\authormark{1}
Xiaojun Xu,\authormark{1,2,3}
and Chaofan Zhang\authormark{1,$\ddagger$}
}

\address{\authormark{1}College of Advanced Interdisciplinary Studies, National University of Defense Technology, Changsha 410073, China\\
\authormark{2}Nanhu Laser Laboratory, National University of Defense Technology, Changsha 410073, China\\
\authormark{3}Hunan Provincial Key Laboratory of High Energy Laser Technology, National University of Defense Technology, Changsha 410073, China\\
\authormark{4}Qingdao Innovation and Development Center, Harbin Engineering University, Qingdao, 266500, China\\
\authormark{$\dag$}The authors contributed equally to
this work.}

\email{\authormark{*}mingl24@nudt.edu.cn\\
\authormark{$\ddagger$}c.zhang@nudt.edu.cn} 


\begin{abstract*}
Photonic-crystal surface-emitting lasers (PCSELs)
have garnered significant attention due to their ability to generate laser beams
with ultra high power and low divergence.
This is because they support high power single mode lasing with volumes orders of magnitude larger
than those of conventional semiconductor lasers.
The finite lateral size in a PCSEL is a primary factor limiting its lasing mode volume
and consequently, its output power.
We demonstrate that the scaling relation between the total cavity loss
$\alpha=\alpha_\perp + \alpha_\parallel$
and the device size $L$ is such that the surface radiation loss scales as $\alpha_{\perp} \sim O(L^{-2})$,
while the edge radiation loss $\alpha_{\parallel} \sim O(L^{-3})$.
Both scaling relations can be explained by the second order expansions of
the complex frequency $\omega(k)$ of the band diagram.
\modify{Our results provide a guideline for estimating the size effect when designing large area PCSEL-like devices.
}

\end{abstract*}

\section{Introduction}
 Over the past two decades, photonic-crystal surface-emitting lasers (PCSELs)
 have attracted increasing attention due to their potential for achieving
 single mode lasing with ultra-high power, compact size,
 low-cost and long lifetime\cite{RN14,RN39,RN40,RN84,RN64,RN123,RN79,RN50,RN163,RN450}.
 Such devices have long been anticipated for various applications,
 including long range optical communication\cite{RN424,RN432,RN157,RN438},
 light detection and ranging (LiDAR)\cite{RN150,RN237},
 laser processing for smart manufacturing\cite{RN79,RN441}, and military applications.
 Compared to conventional semiconductor lasers, the photonic crystal (PhC) slab layer in a PCSEL
 modulates the optical field so that the effective mode volume can be orders of magnitude larger,
 while maintaining relatively robust single mode operation.
 With the energy density limited by current injection, thermal management\cite{RN235}
 and material catastrophic optical damage (COD)\cite{RN457},
 the mode volume, and consequently the effective device size, is critical for realizing high power PCSELs.

\begin{figure}
 	\centering
 	\includegraphics[width=1\linewidth]{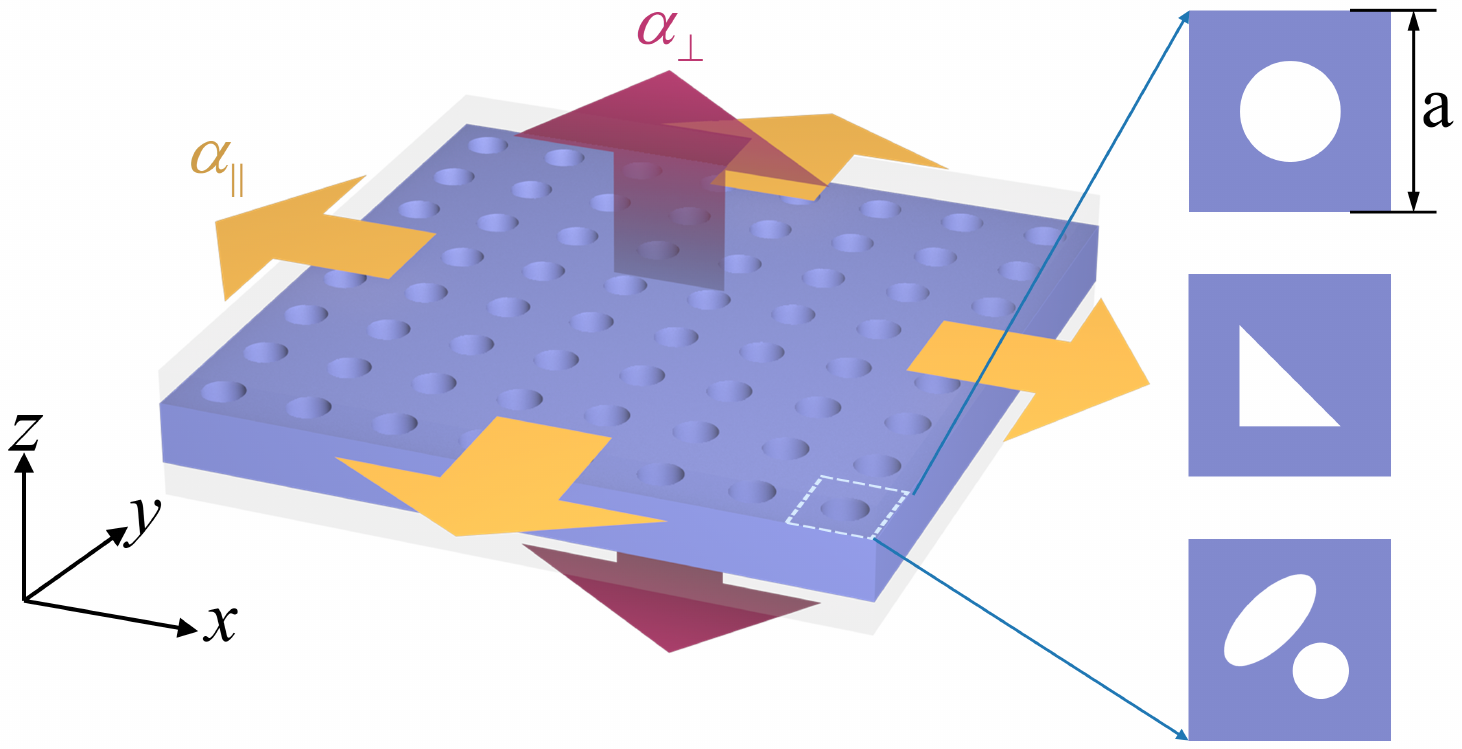}
 	\caption{The schematic of the structure of a PCSEL.
 		The red and the orange arrows denote the surface $\alpha_\perp$
 		and the edge radiation loss $\alpha_\parallel$, respectively.
 		The gray area represents the non-PhC layers in the device,
 		that may include cladding layers, gain media layers, etc.
 		The three insets show the simulated unit cell structure with circular, isosceles right-triangle,
 		and double-lattice holes.
 		The double-lattice unit cell consists of both elliptical and circular holes.}
 	\label{fig:PC_IMAG}
 \end{figure}

 The core structure of a typical photonic crystal surface emitting device is illustrated in Fig.~\ref{fig:PC_IMAG}.
 A PhC slab is sandwiched between cladding layers.
 In the direction perpendicular to the wafer plane ($z$-axis), a PCSEL is essentially a slab waveguide,
 with the optical mode confined around the PhC layer and the gain region,
 but within the wafer plane ($x$-$y$ plane), the ``laser cavity'' in a PCSEL is exceptional:
 although the optical field extends laterally within the wafer plane,
 it is not necessarily confined by reflective boundaries at the device edges.
 Instead, the optical confinement arises by itself near the $\Gamma$-point of the photonic band,
 where the group velocity $v_g$ approaches zero.
 Such resonance mode radiates both to the surface-emitting wave (denoted $\alpha_\perp$ in the figure)
 and to the device edge (denoted $\alpha_\parallel$)\cite{RN50,RN54}.
 For typical PCSEL applications, the surface radiation constitutes the desired output,
 whereas the edge radiation contributes to optical losses.
 The total cavity loss, which includes both the surface and the edge radiation components,
 governs the lasing gain threshold and the optical efficiency.

Understanding the relationship of the surface ($\alpha_\perp$)
and the edge radiation ($\alpha_\parallel$) losses
with respect to the device size $L$ is essential for estimating the device performance.
As the device size increases, the surface radiation loss asymptotically approaches a constant value,
especially zero for bound states in the continuum (BIC)\cite{RN54},
whereas the edge radiation loss approaches zero\cite{RN451,RN419,RN224}.
Most prior studies on the modeling and designing of PCSEL structures
have focused primarily on surface radiation\cite{RN151,changPhotonicCrystalSurface2024,Lang:25}.
It is intuitive but implicitly assumed that the boundary effect and edge radiation should
reduce faster than that of the surface radiation and should be negligible for large area PCSELs.
Although the finite size effect of a PCSEL device has been studied in some previous works\cite{RN54},
the criteria for this assumption have never been examined in any publication to the best of our knowledge.
\modify{The edge radiation loss has been shown to decrease with increasing device size theoretically and experimentally\cite{RN434}.
Recent studies\cite{RN451,RN480,RN481,RN455} observed disagreement of the scaling of edge radiation between different models and numerical simulations.
Specifically, their theoretical model (kSWLE) reported an $O(L^{-2})$ scaling for edge radiation, which is of the same order as the surface radiation.
While these results offer valuable insight, a more detailed treatment of edge radiation scaling remains to be explored.
}

In this work, we refine the established three-dimensional coupled-wave theory (3D-CWT)\cite{supple}
and investigate the radiation loss as a function of the lateral device size.
The asymptotic scaling relation between the radiation loss and the device size is concluded
as $\alpha_\perp\sim O(L^{-2})$ and $\alpha_\parallel\sim O(L^{-3})$.
We further clarify phenomenologically the underlying physical mechanism of this scaling behavior,
where the scaling of the surface radiation $\alpha_\perp$ can be explained by
the mode distribution in $k$ space, and the scaling of the edge radiation $\alpha_\parallel$
\modify{ can be interpreted in terms of wave‑packet spreading.}
\modify{Our findings establish a method for estimating the device size at which the edge radiation becomes negligible, and provide guidelines that, when combined with experimental calibration, can inform the design and scaling of PCSELs of various sizes.}
For applications of PCSELs of medium size\cite{RN399,RN150},
this result helps with the system design and determining the optimum size of the PCSEL chip.
More broadly, the analysis elucidates a general connection between photonic band dispersion
and confinement in open optical Bloch-wave systems,
offering insight into finite-size effects in extended photonic resonators
beyond the specific context of PCSEL devices.

\section{Method}
\subsection{Structural Geometry}
As shown in Fig.~\ref{fig:PC_IMAG},
the simulated structure consists of a PhC layer based on a square lattice,
with artificial unit cells of circular, isosceles right-triangular\cite{RN53},
and double-lattice geometries\cite{RN123}.
The modeling is generally scale-invariant, and so are most of our results,
but for the convenience of comparing the results with previous works and
for facilitating experimental parameter estimations,
a lattice constant of $a = \SI{295}{nm}$ is adopted for assigning physical dimensions to the numbers\cite{RN53}.
The filling factor $f$, defined as the area ratio of the hole in a unit cell,
is $0.16$ throughout this work unless otherwise specified.
The geometry and the optical parameters for the layers used in our simulation are listed in
Table.~\ref{tab:table1}\cite{RN53}.
The size of the circular and the triangular holes in the unit cell is determined by the filling factor $f=0.16$.
The double-lattice unit cell comprises an elliptical and a circular element,
with a center-to-center distance of $0.257a$ between them.
The ellipse has an aspect ratio of $0.45$,
corresponding to the Structure III reported in Ref.~\cite{RN123}.
In the double-lattice unit cell,
the filling factors are $0.045$ for the elliptical and $0.035$ for the circular, respectively.
Unless otherwise specified, all structures in this work adopt the configuration described above.

\begin{table}[htbp]
	\centering
	\caption{Layered structural parameters of the PCSEL\cite{RN53}}
	\label{tab:table1}
	\begin{tabular}{c | c  c}
		\hline\hline
		Layer & Thickness(\si{\micro m}) & Dielectric constant \\
		\hline
		N-clad (\ce{AlGaAs})    & 1.5000  & 11.0224 \\
		Active    & 0.0885 & 12.8603 \\
		PhC    & 0.1180 & 
		$\epsilon_{\text{av}}$ \\
		GaAs   & 0.0590 & 12.7449 \\
		P-clad (\ce{AlGaAs})    & 1.5000 & 11.0224 \\
		\hline\hline
	\end{tabular}
\end{table}

\subsection{Coupled-Wave Theory}
\begin{figure}
	\centering
	\includegraphics[width=0.85\linewidth]{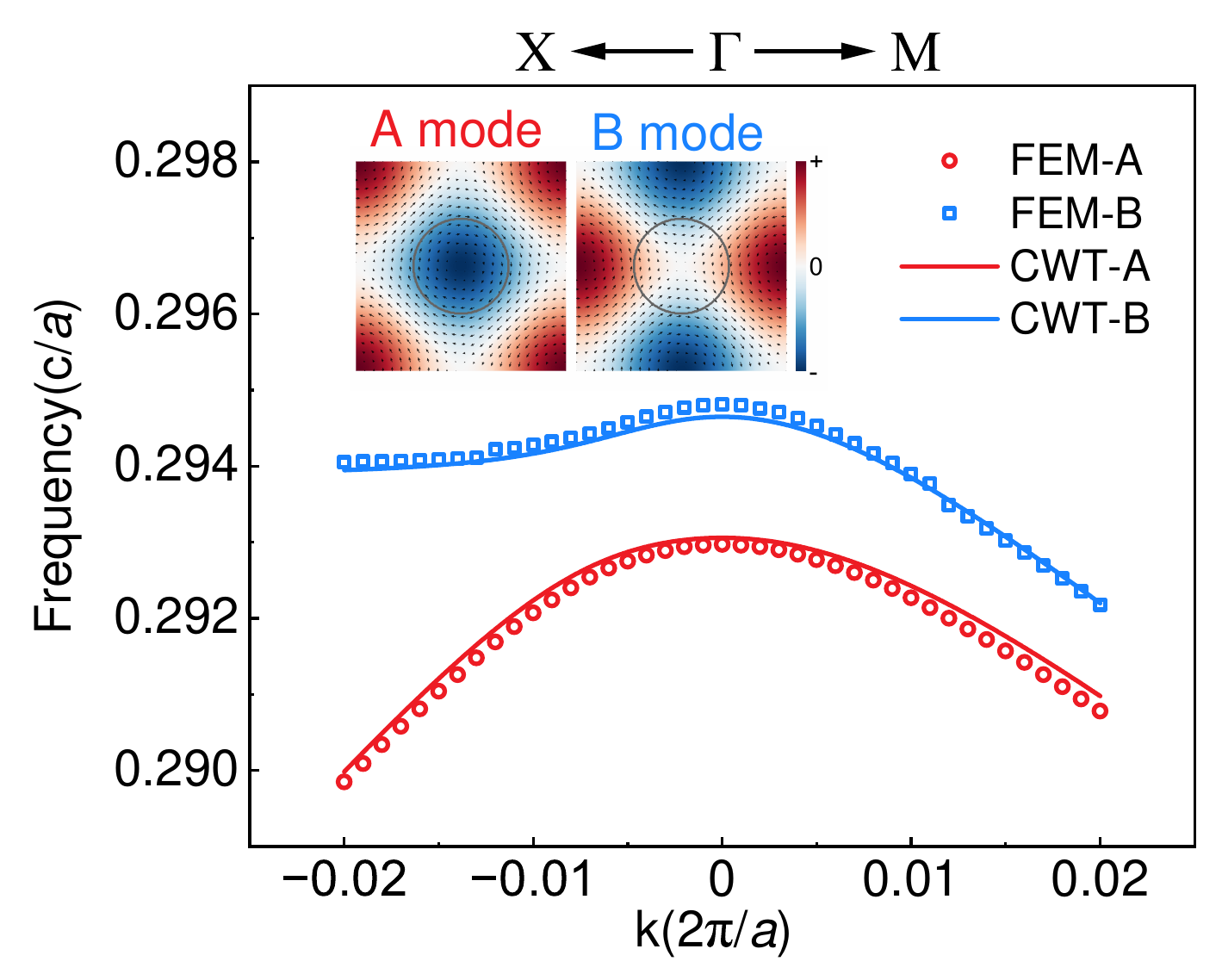}
	\caption{Comparison of the band structure calculated using 3D-CWT (solid lines)
		and Finite Element Modeling (FEM, dotted lines)
		for modes $\mathrm{A}$ and $\mathrm{B}$.
		The insets show the magnitude of the magnetic field (in color)
		and the direction of the electric field (indicated by black arrows)
		within a unit cell for modes $\mathrm{A}$ and $\mathrm{B}$, respectively.
		The plot is calculated for circular unit cell with filling factor $f=0.12$.
	}
	\label{fig:band_structures}
\end{figure}

A naive full numerical electromagnetic solver is not readily feasible
in the case of a PCSEL device simulation,
as the characteristic size of the structure spans approximately five orders of magnitude---from $\SI{E1}{nm}$
(the pattern within a unit cell) to roughly $\SI{E0}{mm}=\SI{E6}{nm}$(the size of the device),
and therefore the number of mesh grid is unacceptably large.
The three-dimensional coupled-wave theory (3D-CWT) provides
a semi-analytical framework to address this challenge\cite{RN54,RN53}.
We revisit the derivation and the interpretation of 3D-CWT, demonstrating that
it is fundamentally a second-order perturbation theory applied to a set of
four-fold degenerate states/modes (section 1 in \cite{supple}).

Along the $z$-axis, the structure can be modeled as a standard slab waveguide,
and its dispersion relation $\omega_\text{slab}(k)$
can be solved using the transfer matrix method\cite{Chilwell:84,Lyu:21}.
Specifically, the average dielectric constant for the PhC layer for TE modes is determined by:
\begin{equation}\label{eq:average_dielectric_constant}
	\epsilon_{\text{av}}=f\epsilon_0+(1-f)\epsilon_{\ce{GaAs}}
\end{equation}
where $\epsilon_0=1$ and $\epsilon_{\ce{GaAs}}=12.7449$ denote the relative dielectric constants of
vacuum (approximately, air) and \ce{GaAs}, respectively.
In the $x$-$y$ plane,
the degenerate basis in 3D-CWT are determined by the Bragg condition for a square lattice:
$\vec k=(2\pi/a, 0), (-2\pi/a, 0), (0, 2\pi/a), (0, -2\pi/a)$.
The corresponding four Fourier components of the electric field are denoted as
$R_x$, $S_x$, $R_y$ and $S_y$, respectively.
The frequency response of a PCSEL is described by the following eigenvalue equation:
\begin{equation}\label{eq:finite_equation}
	\left(\delta+\mi\frac{\alpha}{2}\right)
	\begin{pmatrix}
		R_x \\
		S_x \\
		R_y \\
		S_y
	\end{pmatrix}
	=C
	\begin{pmatrix}
		R_x \\
		S_x \\
		R_y \\
		S_y
	\end{pmatrix}+\mi
	\begin{pmatrix}
		\partial R_x/\partial x \\
		-\partial S_x/\partial x \\
		\partial R_y/\partial y \\
		-\partial S_y/\partial y
	\end{pmatrix}
\end{equation}
where
\begin{equation}\label{eq:C_equation}
	C=\left(C_{\text{1D}}+C_{\text{2D}}+C_{\text{rad}}\right)
\end{equation}
and $\delta = \beta-\beta_{0}$ represents the detuning of the effective wavenumber from the Bragg condition,
$\alpha$ is the total optical power loss,
and $C_\text{1D}$, $C_\text{2D}$ and $C_\text{rad}$ are the matrices corresponding to the coupling between
four basic modes, coupling to higher-order modes and coupling to surface radiation, respectively
\cite{supple,RN50}.
The last term in Eq.~(\ref{eq:finite_equation}) is
the spatial variation of the envelope function of the field within a finite-sized cavity,
capturing the influence of the finite-size effect.
The out-propagating boundary condition
$R_x|_{x=0}=0$, $S_x|_{x=L}=0$, $R_y|_{y=0}=0$, $S_y|_{y=L}=0$ is applied,
which corresponds to no reflection at the boundaries\cite{RN54}.

The complex frequency on the photonic band can be obtained
by extending Eq.~(\ref{eq:finite_equation}) with $\mi\nabla = \vec k$ substitution:
\begin{equation}\label{eq:finite_equation_with_k}
	\left(\delta+\mi\frac{\alpha}{2}\right)
	\begin{pmatrix}
		R_x \\
		S_x \\
		R_y \\
		S_y
	\end{pmatrix}
	=C
	\begin{pmatrix}
		R_x \\
		S_x \\
		R_y \\
		S_y
	\end{pmatrix}+
	\begin{pmatrix}
		k_x R_x\\
		-k_x S_x \\
		k_y R_y \\
		-k_y S_y
	\end{pmatrix}
\end{equation}

\modify{
Notice that Eq.~(\ref{eq:finite_equation_with_k}) implicitly assumes an infinite size or periodic boundary condition, where the wavevector
$k$ is conserved and the eigenmode is also an eigenmode of $k$.
Physically, Eq.~(\ref{eq:finite_equation_with_k}) describes the infinite-periodic Bloch-wave near the $\Gamma$ point.
It captures the branch dispersion and the surface-radiation loss originating from the periodic perturbation, but it does not include the boundary-induced lateral leakage that arises from the finite size of the device.
}The band diagram, which is the real part of the dispersion relation,
can be expressed as $\real \omega(\vec {k}) = c(\beta_{0}+\delta)/n_\text{eff}$
based on Eq.~(\ref{eq:finite_equation_with_k}).
The effective refractive index $n_\text{eff}$ is
solved from the guided mode in the slab waveguide by the transfer matrix method,
$\omega_\text{slab}(\beta_0)=c\beta_0/n_\text{eff}$,
and $c$ denotes the speed of light in vacuum.

The four eigenmodes obtained from Eq.~(\ref{eq:finite_equation_with_k})
are labeled $\mathrm{A},\mathrm{B},\mathrm{C}$ and $\mathrm{D}$
in order of increasing frequency.
Modes C and D are typically highly lossy(leaky) and couple strongly to traveling wave solutions.
As a result, they are of limited relevance for our laser cavity formation and
hard to distinguish in the results from finite-element simulations\cite{RN53}.
In the following discussion, we restrict our analysis to modes $\mathrm{A}$ and $\mathrm{B}$.
Eigenmodes from Eq.(~\ref{eq:finite_equation}) can be distinguished and grouped
to these modes.

Fig.~\ref{fig:band_structures} shows the band diagram of the A and
B modes for a circular unit cell with filling factor $f=0.12$,
the result is compared with the finite element modeling (FEM) result.
The excellent agreement between the two approaches validates the reliability of our implementation of the 3D-CWT.

\subsection{Radiation Loss}
\begin{figure}
	\centering
	\includegraphics[width=0.9\linewidth]{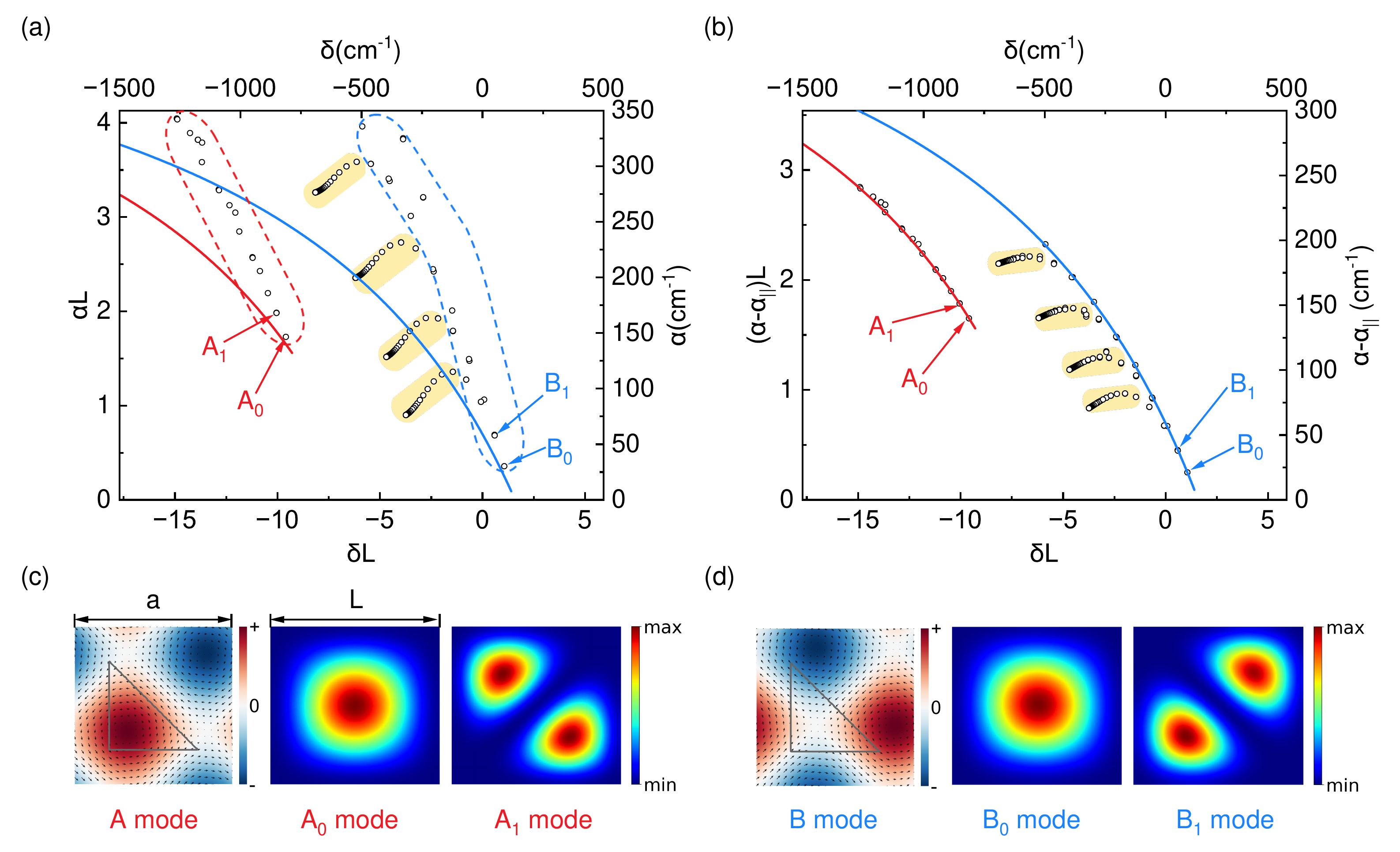}
	\caption{Mode spectrum.
		\textbf{(a)} The mode spectrum of a finite-size PCSEL with $L=400a$.
		Resonance modes $\mathrm{A}$ and $\mathrm{B}$ are marked by red and blue dashed circles, respectively.
		The basic modes $\mathrm{A_0}$ and $\mathrm{B_0}$, as well as the first higher-order modes $\mathrm{A_1}$ and $\mathrm{B_1}$, are indicated with arrows.
		Data points highlighted in yellow are nonphysical artifacts arising from discretizations\cite{RN148}.
		The solid red and blue lines depict the parametric plot of $\delta(k)-\alpha(k)$
		for modes $\mathrm{A}$ and $\mathrm{B}$, obtained via Eq.~(\ref{eq:finite_equation_with_k}).
		\textbf{(b)} The mode spectrum after eliminating edge radiation loss for each mode in (a).
		\textbf{(c)}, \textbf{(d)}
		The magnitude of the magnetic field (in color) and the direction of the electric field (indicated by black arrows)
		within a unit cell for modes $\mathrm{A}$ and $\mathrm{B}$, respectively.
		The field intensity envelopes for modes labeled $\mathrm{A_0}, \mathrm{A_1}, \mathrm{B_0}$ and $\mathrm{B_1}$.
	}
	\label{fig:mode_spectrum}
\end{figure}

The linear operator in Eq.~(\ref{eq:finite_equation}) is non-Hermitian
and therefore naturally yields complex eigenvalues,
whose imaginary parts represent the cavity loss.
This cavity loss arises from optical radiation, and can be further decomposed into:
\begin{equation}
	\alpha=\alpha_{\perp}+\alpha_{\parallel}
\end{equation}
where $\alpha_{\perp}$ is the surface radiation loss and
$\alpha_{\parallel}$ represents the edge radiation loss\cite{RN64},
as illustrated in Fig.~\ref{fig:PC_IMAG}
with red and orange arrows indicating $\alpha_{\perp}$ and $\alpha_{\parallel}$,
respectively.
This model was first introduced in Ref.~\cite{RN54},
which laid an important foundation for subsequent developments.

The detail of the radiation loss is obtained by exploiting the Hermiticity of Eq.~(\ref{eq:C_equation}).
The first two terms, $C_\text{1D}$ and $C_\text{2D}$ are Hermitian.
The cavity loss contributions from $C_\text{rad}$ and the boundary condition yield anti-Hermitian components:
the former gives $\alpha_\perp$, and the latter contributes $\alpha_\parallel$.
Consequently, the imaginary part of the eigenvalue can be expressed as (section 2 in \cite{supple}):
\begin{align}
		&\alpha_{\perp}=2\kappa_{\nu,i}\int_{0}^{L}\int_{0}^{L}
		\left(\left|\xi_{-1,0}R_{x}+\xi_{1,0}S_{x}\right|^{2}+\left|\xi_{0,-1}R_{y}+\xi_{0,1}S_{y}\right|^{2}\right)
		\dif x\dif y
		\label{eq:perp}
		\\
		&\alpha_{\parallel}=
		\int_{0}^{L}\left(\left|R_{ex}(y)\right|^{2}+\left|S_{ex}(y)\right|^{2}\right)\dif y
		+\int_{0}^{L}\left(\left|R_{ey}(x)\right|^{2}+\left|S_{ey}(x)\right|^{2}\right)\dif x
		\label{eq:parallel}
		\\
		&R_{ex}=R_x|_{x=L},\quad S_{ex}=S_x|_{x=0},\quad R_{ey}=R_y|_{y=L},\quad S_{ey}=S_y|_{y=0}
		\label{eq:Rex_Rey_Sex_Sey}
\end{align}
where the coefficient $\kappa_{\nu,i}$ is closely associated with the out-of-plane coupling,
and $\xi_{ij}$ are the Fourier component of the relative dielectric constant distribution
of the PhC layer. Both value are further explained in \cite{supple}.
The first term in Eq.~(\ref{eq:parallel}) can be interpreted as the radiation to the $x=0, L$ boundaries,
denoted by $\alpha_{\parallel x}$,
and the second term $\alpha_{\parallel y}$ corresponds to radiation at $y=0, L$.
An expression analogous to Eq.~(\ref{eq:parallel}) can also be derived using the time-averaged Poynting vector formalism,
differing only by a numerical factor.
This further validates the validity of Eqs.~(\ref{eq:perp}) and (\ref{eq:parallel});
further details are provided in section 3 in \cite{supple}.

We examine the mode spectrum of a PCSEL device with an isosceles right-triangular unit cell and size $L=400a$.
The result is qualitatively the same for different sizes $L$ that are not too small.
\modify{For very small $L$, the slow-varying envelope approximation breaks down and the mode mixing increases}
(section 2 in \cite{supple}).
The total optical power loss $\alpha$ is plotted against the Bragg condition detuning $\delta$
in Fig.~\ref{fig:mode_spectrum}(a) as circles for each mode.
The red and blue dashed lines in Fig.~\ref{fig:mode_spectrum}(a)
enclose the mode spectra for modes $\mathrm{A}$ and $\mathrm{B}$, respectively.
This result is consistent with that reported in Ref.~\cite{RN148}.
In previous literature\cite{RN52,RN54},
mode spectrum plots have typically been obtained from finite-sized devices in real space.
However, the imaginary part of the band diagram should inherently contain information
about the optical loss for infinite-size devices—specifically, the surface radiation loss $\alpha_\perp$.
To confirm the interpretation,
$\alpha_\perp$ versus $\delta$ is plotted in Fig.~\ref{fig:mode_spectrum}(b) as circles for each mode.
The solid red and blue lines in Figs.~\ref{fig:mode_spectrum}(a) and (b)
represent the mode spectrum solved from Eq.~(\ref{eq:finite_equation_with_k}).
These lines align perfectly with the low order modes in Fig.~\ref{fig:mode_spectrum}(b),
confirming not only the validity of the above interpretation but also the reliability of the proposed radiation loss model.
Furthermore, the loss component $\alpha_{\perp}$ can be understood from a band-structure perspective,
as will be demonstrated in the following section.

Additionally, Fig.~\ref{fig:mode_spectrum}(c) shows the magnetic field distribution (color map)
and electric field polarization (indicated by arrows) of modes $\mathrm{A}$ within a single unit cell,
as well as the field intensity envelopes for the fundamental mode
$\mathrm{A_0}$ and the first higher-order mode $\mathrm{A_1}$.
Similar to Fig.~\ref{fig:mode_spectrum}(c), Fig.~\ref{fig:mode_spectrum}(d) depicts the characteristics of mode B.

\section{Result}
In this section, we employ the model developed in the previous sections to
investigate how the radiation loss scales with device size $L$ in square PCSELs,
and discuss the underlying physical mechanisms of the observed relations.
\subsection{Numerical Result}
\begin{figure}
	\centering
	\includegraphics[width=1\linewidth]{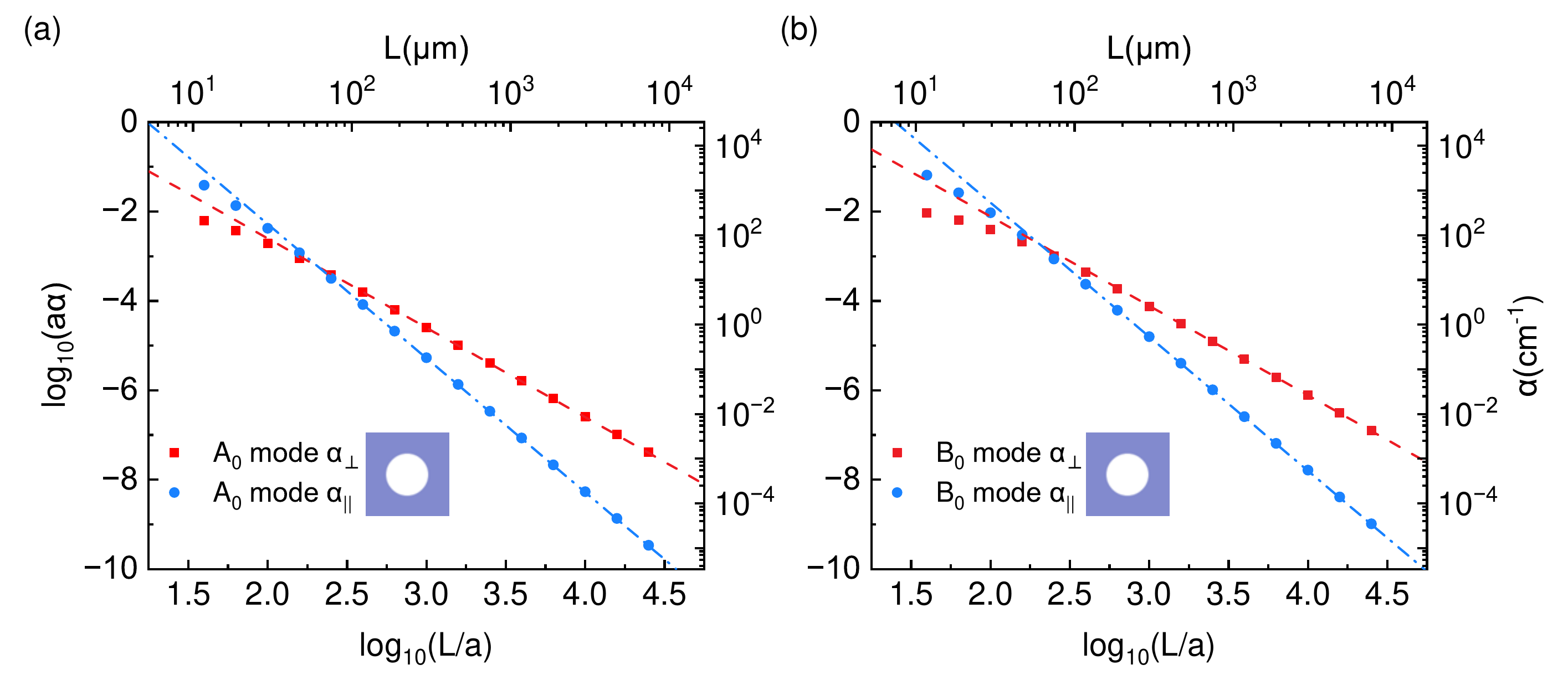}
	\caption{The surface radiation loss ($\alpha_{\perp}$) and the edge radiation loss ($\alpha_{\parallel}$)
		for PCSEL devices with circular unit cells (the inset) in square lattices of varying device size $L$.
		\textbf{(a)} $\mathrm{A_0}$ mode.
		\textbf{(b)} $\mathrm{B_0}$ mode.
		The dashed lines are the linear fits for data points with $L>1000a$.
	}
	\label{fig:rad_circle}
\end{figure}

\begin{figure}
	\centering
	\includegraphics[width=0.9\linewidth]{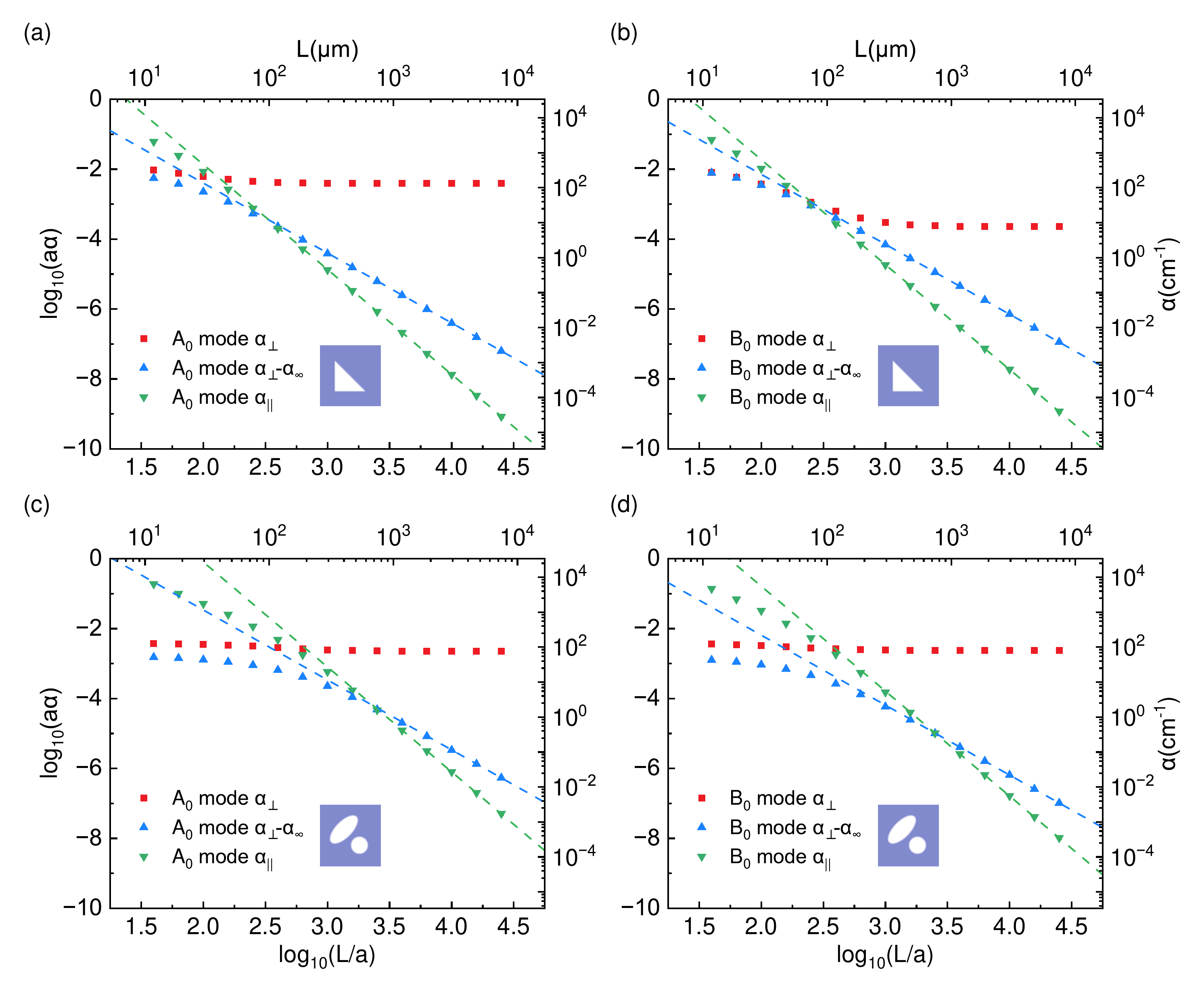}
	\caption{The radiation loss of the $\mathrm{A_0}$ and $\mathrm{B_0}$ modes for PCSELs with different unit cell structure.
		\textbf{(a)}, \textbf{(b)} isosceles right-triangular and
		\textbf{(c)}, \textbf{(d)} double-lattice, see the inset.
		$\alpha_\infty$ is the surface radiation for $L\to\infty$ of the same structure.
	}
	\label{fig:rad_IRT_DL}
\end{figure}

\begin{figure}
	\centering
	\includegraphics[width=0.9\linewidth]{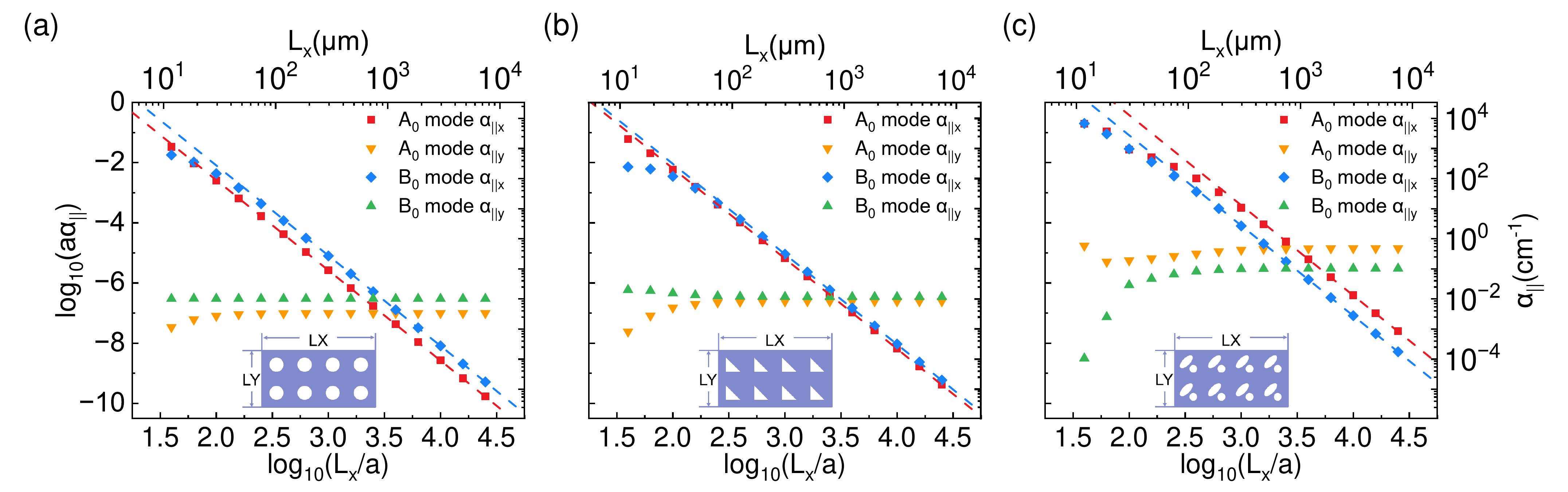}
	\caption{The edge radiation loss for rectangular devices.
		The inset shows a schematic of the rectangular devices,
		where $L_{y}$ is held constant at $3000a$, and $L_x$ is varied;
		$\alpha_{\parallel x}$ represents the radiation leakage from the left and the right edges,
		and $\alpha_{\parallel y}$ denotes the radiation leakage from the top and bottom edges.
		\textbf{(a)} Circular unit cell
		\textbf{(b)} Isosceles right triangular unit cell
		\textbf{(c)} Double‑lattice unit cell.
	}
	\label{fig:rectangular_rad}
\end{figure}

Figs.~\ref{fig:rad_circle} (a) and (b) illustrate the asymptotic dependence of $\alpha_{\perp}$ and $\alpha_{\parallel}$
on device size $L$ for the $\mathrm{A_0}$ and $\mathrm{B_0}$ modes, respectively, in square-lattice PCSELs with circular unit cells.
The dashed lines are linear fits for the data points of large $L$ in the log scale,
with the slope being $-3$ for $\alpha_\parallel$ and $-2$ for $\alpha_\perp$.
This result indicates that for both modes $\mathrm{A_0}$ and $\mathrm{B_0}$,
$\alpha_{\perp}\sim O(L^{-2})$ and $\alpha_{\parallel}\sim O(L^{-3})$.

A PhC slab with a circle unit cell exhibits almost zero surface radiation due to its high symmetry
and the presence of a bound state in the continuum (BIC) at the $\Gamma$ point\cite{RN53,RN400}.
For typical PCSEL applications, however, devices that radiate at the $\Gamma$ point are preferred.
Some such structures are shown in Fig.~\ref{fig:rad_IRT_DL}.
$\alpha_\perp$ converges to a constant $\alpha_\infty$ for $L\to\infty$.
The blue dots are obtained by subtracting $\alpha_{\infty}$
of the corresponding infinitely large PCSELs from the surface radiation loss $\alpha_\perp$.
$\alpha_\infty$ is also the $\Gamma$ point loss, and a natural result solved from
Eq.~(\ref{eq:finite_equation}) without the last term of the spatial dependence.
Further details are provided in \cite{supple}.
This is also how we calculate $\alpha_\infty$ in practice,
and the limit $\lim_{L\to\infty}\alpha_\perp = \alpha_\infty = \alpha (k=0)$
is verified in Fig.~\ref{fig:rad_IRT_DL} as $\alpha_\perp - \alpha_\infty$
goes to zero with large $L$.
These results show that:
\begin{align}
	\alpha_\perp-\alpha_{\infty} &\sim O(L^{-2}) \label{eq:alpha_perb-alpha} \\
	\alpha_{\parallel} &\sim O(L^{-3}) \label{eq:alpha_paralle}
\end{align}
Specifically, for PCSELs supporting BIC modes at the $\Gamma$ point, $\alpha_\infty = 0$.

\begin{figure}
	\centering
	\includegraphics[width=0.9\linewidth]{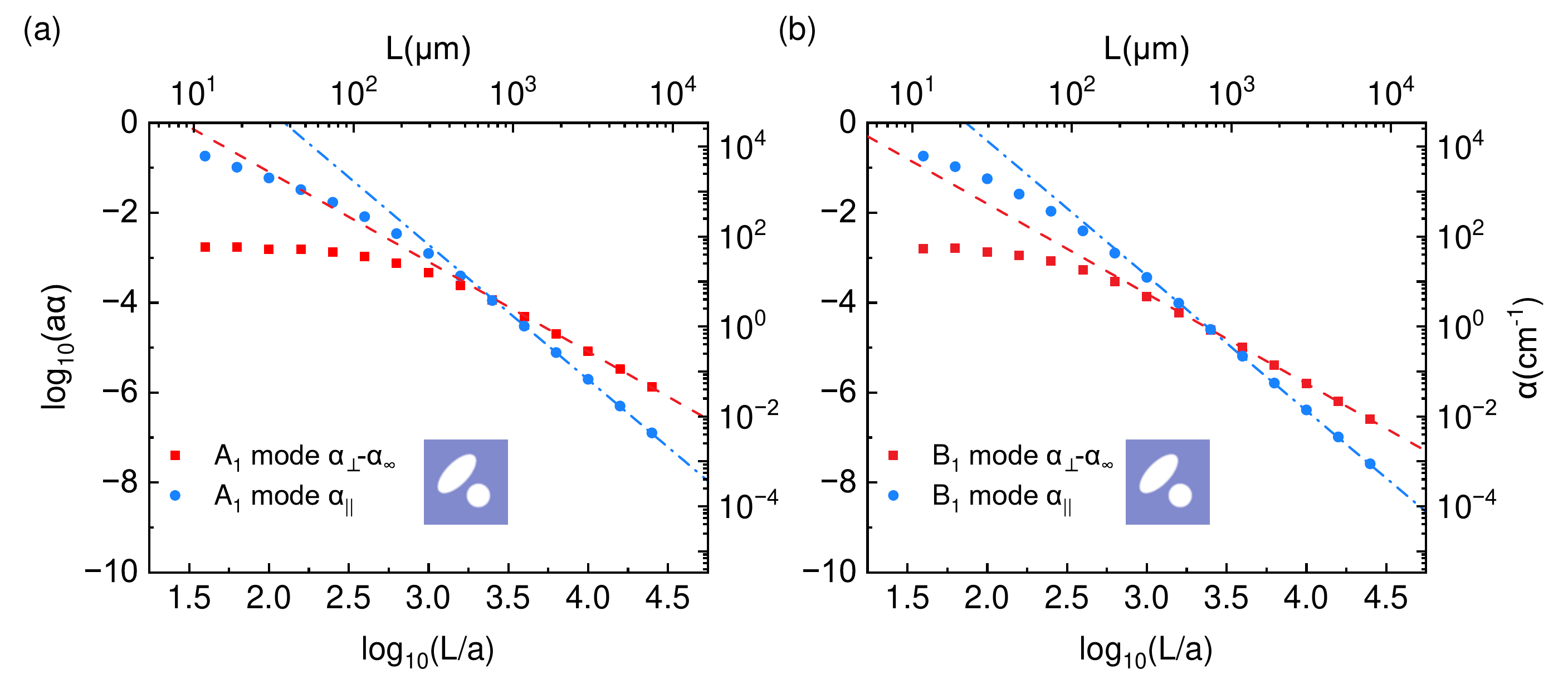}
	\caption{The radiation loss of the first higher-order modes:
		(a)$\mathrm{A_1}$ and (b)$\mathrm{B_1}$, in PCSELs with a double-lattice unit cell.
	}
	\label{fig:first_higher}
\end{figure}

To further investigate the dependence of the results on the device geometry,
finite-size rectangular devices are studied in Fig.~\ref{fig:rectangular_rad},
with fixed width along $y$-axis $L_y=3000a$ and varying $L_{x}$.
$\alpha_{\parallel x}$ denotes the edge radiation loss along the $x$-direction
(emitted from the edges of length $L_y$),
while $\alpha_{\parallel y}$ refers to the loss along the $y$-direction
(emitted from the edges of length $L_x$).
The result shows that the edge radiation loss $\alpha_{\parallel x}$ and $\alpha_{\parallel y}$
scales independently with $L_x$ and $L_y$, respectively,
i.e. $\alpha_{\parallel x}\sim O(L_x^{-3})$ and $\alpha_{\parallel y}\sim O(L_y^{-3})$.
Furthermore, we investigate the scaling behavior of higher-order modes.
Figs.~\ref{fig:first_higher} (a) and (b) show the variation of radiation loss with device length $L$ for the first higher-order modes $(\mathrm{A_1})$ and $(\mathrm{B_1})$, respectively.
The higher order modes exhibit the same behavior of the fundamental modes presented earlier.

\subsection{Physics Interpretation}
\begin{figure}
	\centering
	\includegraphics[width=0.9\linewidth]{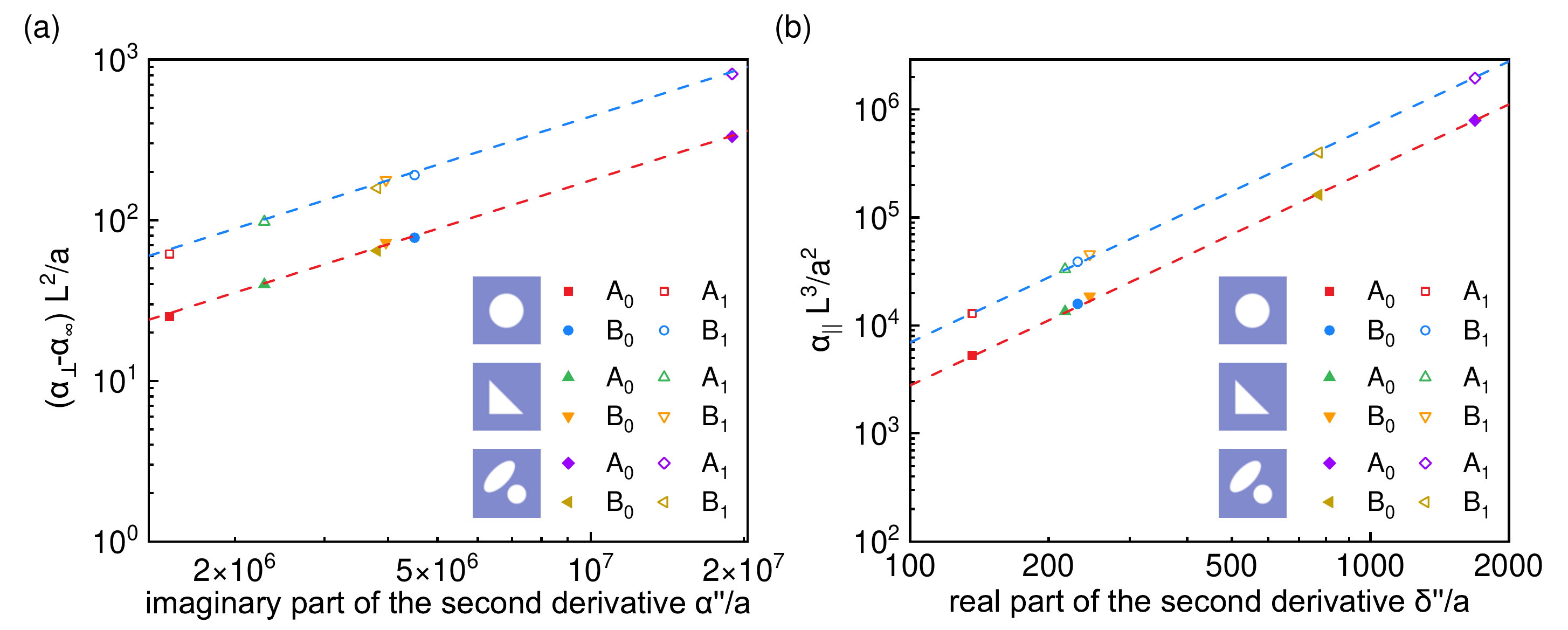}
	\caption{The relationship between surface radiation loss $(\alpha_\perp)$ (\textbf{a}),
		edge radiation loss $(\alpha_{\parallel})$ (\textbf{b}) and
		the second derivative of the dispersion relation $\omega(k)$.
		The slope of dashed lines is $1$ in (a) and $2$ in (b).
	}
	\label{fig:LLL_evident}
\end{figure}

Both scaling relations Eqs.~(\ref{eq:alpha_perb-alpha}) and (\ref{eq:alpha_paralle})
can be explained semi-quantitatively by the band diagram.
The Taylor expansion of $\omega(k)$ around the $\Gamma$ point is given by:
\begin{align}
	&\frac{n_\text{eff}}{c}\omega(k)-\frac{2\pi}{a}=
	\delta(k)+\mi \frac{\alpha(k)}{2}
	\label{eq:omega_k_taylor} \\
	&\delta(k)=\delta_\infty+\frac{\delta''}{2} k^2+O(k^4)
	\label{eq:omega_k_taylor1} \\
	&\alpha(k)=\alpha_\infty+\frac{\alpha''}{2}k^2+O(k^4)
	\label{eq:omega_k_taylor2}
\end{align}
The terms of odd power are zero because of the PT-symmetry of the PhC structure, where $\omega(k)=\omega(-k)$.

The scaling of the surface radiation, Eq.~(\ref{eq:alpha_perb-alpha}),
is a direct result of the imaginary part in Eq.~(\ref{eq:omega_k_taylor2}).
For a wave-packet of finite size $L$, the basic mode distribution in the $k$-space is $|k|\sim\pi/L$.
Substituting this into Eq.~(\ref{eq:omega_k_taylor2}) directly yields Eq.~(\ref{eq:alpha_perb-alpha}), i.e.
\begin{equation}
	\alpha_\perp - \alpha_\infty \propto \frac{\alpha''}{L^2}
	\label{eq:alpha_perp_to_dispersion}
\end{equation}
Specifically in the case of a circular unit cell (BIC at $\Gamma$), where $\alpha_{\infty}=0$,
the surface radiation $\alpha_\perp = O(L^{-2})$ is the result of the mode expanding away from the BIC.
This result is the same as what is recently presented in Ref.~\cite{RN451}.

The scaling of the edge radiation, Eq.~(\ref{eq:alpha_paralle}), is the result of wave-packet spreading.
As a toy model,
the time evolution of the wave-packet size $\sigma_r(t)$ in a Gaussian packet is examined in
section 4 in \cite{supple}.
\modify{For a realistic wave-packet, the first‑order spreading velocity at $t=0$ vanishes; only second‑order spreading exists.
In the small$-t$ limit, the spreading rate scales as}
\begin{equation}
	\left.\frac{\partial\sigma_r(t)}{\partial t}\right|_{t\to 0}
	\propto\left.\frac{\delta''^2t}{{\sigma_0}^2\sigma_r(t)}\right|_{t\to 0}
	\propto \frac{\delta''^2}{L^3}
\end{equation}
where $\sigma_0 = \sigma_r(t=0)\sim L$ is roughly the device size.
The expansion of the wave-packet results in energy leak from the device region.
The energy loss of a mode is the result of the mode leaking from the edge of the device,
and
\modify{can be qualitatively related to the expansion of the wave‑packet (as a heuristic argument),}
so
\begin{align}
	\alpha_\parallel
	\sim \left. \frac{\partial \mathcal E(t)}{\partial t} \right|_{t \to 0}
	&\propto \left. \frac{\partial \sigma_r(t)}{\partial t} \right|_{t \to 0}
	\propto \frac{\delta''^2}{L^3}
	\label{eq:alpha_parallel_to_dispersion}
\end{align}

Eqs.~(\ref{eq:alpha_perp_to_dispersion}) and (\ref{eq:alpha_parallel_to_dispersion})
are verified in Fig.~\ref{fig:LLL_evident}(b) for the relationship between
the loss scaling and the second order expansion coefficient of the dispersion relation.
From the numerical result we can see that the scaling relation is indeed
consistent with our physics interpretation above,
and across different unit cell structures and different bands.
Higher order modes ($\mathrm{A_1}$ and $\mathrm{B_1}$) are also presented in the figure,
showing that they obey the same relation, with a different constant factor.
This is because there is a larger $k$ of higher mode indexing in a 2D system.
\modify{This result suggests that the finite size effect can be understood largely from the parameters of the band structure, capturing the dominant trends regardless of the detailed shape and field distribution within a unit cell.}

\modify{Additionally, as can be seen in Figs.~\ref{fig:rad_circle}-\ref{fig:first_higher}, the simulation results deviate from the asymptotic curves when $L$ is small.
We attribute this deviation mostly to the non-parabolic effect of the band.
The coefficients of the quadratic and the quartic terms of $\omega(k)$ have opposite signs, making the actual loss smaller than the parabolic approximation.
A detailed discussion can be found in Supplementary Information Section 5\cite{supple}.}

\section{Conclusion}
In summary, the scaling relation for the surface and the edge radiation loss in finite-size PCSELs is uncovered.
The total radiation loss of a PCSEL device decomposes into an edge component $\alpha_\parallel$
which scales as $\alpha_\parallel = O(L^{-3})$ and a surface component $\alpha_\perp$ which scales as
$\alpha_\perp = \alpha_\infty + O(L^{-2})$.
Both results can be explained and estimated by the band diagram.
The scaling behavior of the surface radiation loss ($\alpha_\perp$)
arises from the mode distribution in $k$-space,
which comprises the radiation contribution at the $\Gamma$ point $\alpha_\infty$
an additional loss term stemming from off- $\Gamma$ components that scale as ($O(k^2)\sim O(L^{-2})$).
In contrast, the edge radiation loss ($\alpha_{\parallel}$) is governed by wave-packet spreading.
The scaling law is further validated across diverse unit cell geometries,
demonstrating their robustness beyond specific lattice designs.
Our work provides a semi-quantitative and physically insightful theory for estimating
radiation loss in PCSELs of arbitrary size and shape,
and establishes the relation between the finite size effect and the band diagram,
offering a solid foundation for the rational design of high-efficiency, scalable PCSEL devices.
Beyond their immediate relevance to PCSEL design,
the scaling relations identified in this work highlight a general connection between
finite-size radiation losses and the local curvature (the second-order expansion)
of complex photonic band dispersions.
The results are largely independent of microscopic unit-cell details and
apply broadly to extended photonic resonators with open boundaries.
This perspective provides a unified framework for estimating confinement and
loss in large-area photonic devices and may be applicable to other systems
such as photonic-crystal resonators,
and surface-emitting structures based on bound states in the continuum,
where mode leakage is governed by band dispersion rather than physical mirrors.

\begin{backmatter}
\bmsection{Funding}
Innovation Research Foundation of the National University of Defense Technology (NUDT)

\bmsection{Acknowledgment}
We would like to acknowledge Prof. Wei Liu for helpful discussions.

\bmsection{Disclosures}
The authors declare no conflicts of interest.

\bmsection{Data availability}
The data that support the findings of this study are available from the corresponding author upon
reasonable request.
\bmsection{Supplemental document}
The supplementary information provides (1) a detailed derivation of the 3D-CWT within the perturbation theory framework; (2) analysis of the anti-Hermitian component in 3D-CWT and the associated mode radiation loss; (3) the edge radiation and the Poynting vector; and (4) investigation of wave-packet spreading for a Gaussian wave-packet. See Supplement 1 for supporting material.

\end{backmatter}
\bibliography{pcsel}

\end{document}


\section{3D-CWT in the perturbation theory framework}
The 3D-CWT model, originally developed and subsequently refined by Liang et al.\cite{RN53,RN54,RN227},
is a well-established framework for simulating photonic crystal surface-emitting lasers (PCSELs),
that agrees well with typical experiments.
For simplicity in the following derivation, we choose the dimensionalization with $c=1$.
The wave equation forms of Maxwell equation yields:
\begin{equation}\label{eq:Electric_field_wave}
	-\nabla\times\nabla\times \vec E(r)
	= \nabla^2\vec E - \nabla(\nabla\cdot\vec E)
	= -\omega^2\epsilon(r)\vec E(r)
\end{equation}
We note the operator in the equation $\mathcal D$ and $\mathcal E$,
and under the real space representation $\mathcal D = -\nabla\times\nabla\times$,
$\mathcal E = \epsilon(r)$.
This is a generalized eigenvalue problem with eigenvalue $\lambda = -\omega^2$,
$\mathcal E$ being the metric operator,
and the inner product is defined as
\begin{equation}
	\left< \vec E_1, \vec E_2\right>
	\equiv \vec E_1^\dagger \vec E_2
	= \int \epsilon(r)\vec E_1^*(r)\cdot \vec E_2(r)\dif V
\end{equation}
In the following we use the normalized field $\phi$ with $\langle\phi, \phi\rangle=1$ and $\vec E=\vec E_0\phi$
for convenience.
$\mathcal E$ can be further decomposed to:
\begin{equation}
	\mathcal E = \epsilon_\text{av}(z) + \mathcal I_\text{PhC}(z)\delta_\epsilon(x, y)
\end{equation}
where
\begin{equation}
	\mathcal I_\text{PhC} = \begin{cases}
		1 & \text{within the PhC layer}\\
		0 & \text{otherwise}
	\end{cases}
\end{equation}
and $\epsilon_\text{av}$ denotes the average dielectric constant of the photonic crystal (PhC) layer,
as explained in the main text, so that
\begin{equation}
	\int\delta_\epsilon(x, y) \dif x\dif y = 0
\end{equation}
i.e. the zero frequency component of the Fourier transform of $\delta_\epsilon$ is zero.
$\delta_\epsilon$ is treated as the perturbation term
($\delta\mathcal E = \mathcal I_\text{PhC}\delta_\epsilon$).

For a system with periodic $\delta_\epsilon$, it would be natural to perform a
Fourier transform in the $x$-$y$ plane.
Consequently, a TE mode can be expressed in the form:
\begin{equation}
	\vec E = \sum_{m, n} \vec E_{m,n}(z) \exp(-\mi \vec \beta_{m,n} \cdot \vec r)
	= \sum_{m, n} \vec E_{m,n}(z) \exp(-\mi m\beta_0 x -\mi n \beta_0 y)
\end{equation}
where $\beta_0 = 2\pi/a$, $m$, $n$ are integers and $E_z = 0$.
Eq.~(\ref{eq:Electric_field_wave}) yields
\begin{equation}
	\left[\frac{\partial^2}{\partial z^2} -
	\begin{pmatrix}
		n^2 & -mn \\
		-mn & m^2
	\end{pmatrix}{\beta_0}^2
	\right]
	\begin{pmatrix}
		E_{x, m, n} \\
		E_{y, m, n}
	\end{pmatrix}
	= -{\omega}^2
	\sum_{m', n'}\xi_{m-m', n-n'}
	\begin{pmatrix}
		E_{x, m', n'} \\
		E_{y, m', n'}
	\end{pmatrix}\label{eq:fourier_xy_real_z}
\end{equation}
\begin{equation}
	\frac{\partial}{\partial z}\left(
	m\beta_0E_{x, m, n} + n\beta_0E_{y, m, n}
	\right) = 0 \label{eq:te_subspace}
\end{equation}
Specifically, $\xi_{0, 0} = \epsilon_\text{av}(z)$, while $\xi_{m, n}$ for $m, n \neq 0$ represent the Fourier
component of $\delta_\epsilon$.
Eq.~(\ref{eq:te_subspace}) defines the subspace for the TE mode solution;
the left-hand side of Eq.~(\ref{eq:fourier_xy_real_z}) corresponds to $\mathcal D\vec E$
represented in the $k$ space.
Specially for the non-perturbed equation ($\delta_\epsilon = 0$) and for
the $(m, n)=(\pm 1, 0)$ or $(0, \pm 1)$, the equation reduces to
\begin{equation}
	\left(
	\frac{\partial^2}{\partial z^2} - \beta_0^2
	\right)\Theta(z)
	= -{\omega_0}^2\epsilon_\text{av}(z)\Theta(z)
	\label{eq:unperturbed_ttm}
\end{equation}
In this context, $\Theta$ represents the normalized scaler mode,
and the electric field is given by $\vec E_{m,n} = E_0\Theta(z)(n\hat x + m\hat y)/\sqrt{m^2 + n^2}$,
which is perpendicular to $\vec \beta_{m, n}$.
This equation is identical to that of a slab waveguide and can be solved analytically using the transfer matrix method.

The perturbative nature of the system is more conveniently expressed in operator formalism than in differential equation form.
We denote the normalized eigenvector (eigenmode) as $\phi$,
with its unperturbed component denoted as $\phi^{(0)}$.
The first-order perturbation yields:
\begin{equation}
	\left(\mathcal D - \lambda^{(0)}\mathcal E\right)\phi^{(1)} = \lambda^{(0)}\delta\mathcal E\phi^{(0)}
	+\lambda^{(1)}\mathcal E\phi^{(0)}
	\label{eq:gen_eigen_perturb_raw}
\end{equation}
Perform $\phi^{(0)\dagger}$ on both sides of the equation
\footnote{It is worth mentioning that, in this context, $\phi^\dagger$ does not represent
	a vector but the operator of non-generalized inner product, i.e.
	$\phi^\dagger\psi = \int \phi^{T*}\psi\dif V$.},
and observe that:
\begin{align}
	&\phi^{(0)\dagger} \mathcal D\phi^{(1)} = (\mathcal D\phi^{(0)})^\dagger\phi^{(1)}
	= \lambda^{(0)}(\mathcal E\phi^{(0)})^\dagger\phi^{(1)}
	= \lambda^{(0)}\phi^{(0)\dagger} \mathcal E\phi^{(1)}
\end{align}
Consequently, the left-hand side (L.H.S.) of Eq.~(\ref{eq:gen_eigen_perturb_raw}) vanishes.
This result also holds for a degenerate perturbation basis.
The right-hand side (R.H.S.) yields:
\begin{equation}
	\lambda^{(1)}
	\phi^{(0)\dagger}_\alpha \mathcal E\phi_\beta^{(0)} =
	-\lambda^{(0)} \phi^{(0)\dagger}_\alpha\delta\mathcal E\phi_\beta^{(0)}
	\label{eq:gen_eigen_first_order_matrix}
\end{equation}
where $\alpha$, $\beta$ are indices labeling the degenerate states.
Since the unperturbed basis $\phi_\alpha^{(0)}$ is chosen to be orthonormal,
we have $\phi^{(0)\dagger}_\alpha \mathcal E\phi_\beta^{(0)} = \delta_{\alpha\beta}$.
The R.H.S of Eq.~(\ref{eq:gen_eigen_first_order_matrix}) represents the matrix elements of $\delta\mathcal E$
under the $\{\phi^{(0)}_\alpha\}$ basis.
Thus, Eq.~(\ref{eq:gen_eigen_first_order_matrix}) implies that the first-order correction
$\lambda^{(1)}$ corresponds to the eigenvalues of the matrix $(\delta\mathcal E)_{\alpha\beta}$,
which is the projection of the operator $\delta\mathcal E$ onto the $\{\phi^{(0)}_\alpha\}$ basis.

For the $\{\phi^{(0)}_\alpha\}$ basis (whose $z$ dependency is defined in Eq.~(\ref{eq:unperturbed_ttm})),
the basis labels are denoted as $\alpha=(m, n)$ and $\beta=(m',n')$.
Due to the definition of $\epsilon_\text{av}$, the diagonal matrix elements for $\alpha = \beta$ vanish.
Furthermore, owing to the orthogonality of the vector field directions,
for all off-diagonal elements $\alpha \neq \beta \in \{(\pm 1, 0), (0, \pm 1)\}$,
only $(\delta\mathcal E)_{(\pm 1, 0), (\mp 1, 0)}$ and
$(\delta\mathcal{E})_{(0,\pm 1), (0, \mp 1)}$ are non-zero. These elements are given by:
\begin{align}
	&(\delta\mathcal E)_{(\pm 1, 0), (\mp 1, 0)}
	=\int\delta_\epsilon\me^{\mp 2\mi\beta_0x}\dif x\dif y\int\mathcal I_\text{PhC}|\Theta(z)|^2\dif z
	=\xi_{\pm 2, 0}\int_\text{PhC}|\Theta(z)|^2\dif z\\
	&(\delta\mathcal{E})_{(0,\pm 1), (0, \mp 1)}
	=\int\delta_\epsilon\me^{\mp 2\mi\beta_0y}\dif x\dif y\int\mathcal I_\text{PhC}|\Theta(z)|^2\dif z
	=\xi_{0, \pm 2}\int_\text{PhC}|\Theta(z)|^2\dif z
	\label{eq:gen_eigen_1st_order_int}
\end{align}

For higher-order perturbation, modes $\psi$ other than the subspace of the $\{\phi^{(0)}_\alpha\}$ basis are involved,
\begin{equation}
	\begin{pmatrix}
		\mathcal D_{11} & 0 \\
		0 & \mathcal D_{22}
	\end{pmatrix}\begin{pmatrix}
		\phi\\\psi
	\end{pmatrix}
	= \lambda
	\begin{pmatrix}
		\mathcal E_{11} & \mathcal E_{12} \\
		\mathcal E_{21} & \mathcal E_{22}
	\end{pmatrix}
	\begin{pmatrix}
		\phi\\\psi
	\end{pmatrix}
	\label{eq:gen_eigen_2nd_order_mat}
\end{equation}
Here, the operator $\mathcal D$ is block diagonal because we choose $\phi$ and $\psi$ as the Fourier basis
in the $x$-$y$ plane, and the perturbation term:
\begin{equation}
	\delta\mathcal E =
	\begin{pmatrix}
		0 & \mathcal E_{12} \\
		\mathcal E_{21} & 0
	\end{pmatrix}
\end{equation}
appears only in the off-diagonal blocks
since $\epsilon_\text{av}(z) = \xi_{0, 0}$. With such labeling,
the equation for the basis $\phi$ reads $\mathcal D_{11}\phi^{(0)}=\lambda^{(0)}\mathcal E_{11}\phi^{(0)}$.

Solving Eq.~(\ref{eq:gen_eigen_2nd_order_mat}) yields:
\begin{align}
	& \psi = \lambda \left(\mathcal D_{22} - \lambda \mathcal E_{22}\right)^{-1}\mathcal E_{21}\phi \\
	& \left(\mathcal D_{11}-\lambda \mathcal E_{11}\right)\phi
	= \lambda^2 \mathcal E_{12} \left(\mathcal D_{22} - \lambda \mathcal E_{22}\right)^{-1}\mathcal E_{21}\phi
	\label{eq:gen_eigen_2nd_order_non_expand}
\end{align}
We expand Eq.~(\ref{eq:gen_eigen_2nd_order_non_expand}) to second order and note that $\mathcal E_{12}$
and $\mathcal E_{21}$ are first-order in the perturbation strength, yielding:
\begin{equation}
	\left(\mathcal D_{11} -\lambda^{(0)}\mathcal E_{11}\right)\phi^{(2)}
	=
	\lambda^{(0)2}\mathcal E_{12} \left(\mathcal D_{22} - \lambda \mathcal E_{22}\right)^{-1}\mathcal E_{21}\phi^{(0)}
	+
	\lambda^{(2)}\mathcal E_{11}\phi^{(0)}
\end{equation}
This equation has the same form as Eq.~(\ref{eq:gen_eigen_perturb_raw}).
Consequently, the perturbation problem reduces to evaluating the matrix elements of the effective operator
$\mathcal E_{12} \left(\mathcal D_{22} - \lambda \mathcal E_{22}\right)^{-1}\mathcal E_{21}$:
\begin{align}
	& \phi^{(0)\dagger}_\alpha
	\mathcal E_{12} \left(\mathcal D_{22} - \lambda \mathcal E_{22}\right)^{-1}\mathcal E_{21}
	\phi^{(0)}_\beta
	= \mathcal I_{\alpha, \nu} \mathcal I_{\beta, \nu}
	\int_\text{PhC} \xi_{\alpha - \nu}\xi_{\beta - \nu}\Theta^*(z)\Psi_\nu(z)\dif z
	\label{eq:gen_eigen_2nd_order_int}
	\\
	& \Psi_\nu(z) = \left[
	\frac{\partial^2}{\partial z^2}
	-({m_\nu}^2 + {n_\nu}^2){\beta_0}^2
	+ \omega_0^2
	\right]^{-1}\mathcal I_\text{PhC}(z)\Theta(z)
\end{align}
where $\nu = (m_\nu, n_\nu)$ represents the Fourier index of the $\psi$ mode.
The $\mathcal I_{\alpha, \nu}$ and $\mathcal I_{\beta, \nu}$ factor comes from the inner
product of the vector field defined by Eq.~(\ref{eq:te_subspace}), and specially
$\mathcal I_{\alpha 0} = 1$ for singularity of  Eq.~(\ref{eq:te_subspace}). 
More explicitly, $\Psi_\nu$ can be expressed as the integral of the Green's function:
\begin{align}
	& \left[
	\frac{\partial^2}{\partial z^2}
	-({m_\nu}^2 + {n_\nu}^2){\beta_0}^2
	+ \omega_0^2
	\right] G_\nu(z, z') = -\delta(z, z') \\
	& \Psi_\nu = \int_\text{PhC} G_\nu(z, z')\Theta(z')\dif z
\end{align}

The final result is obtained by summing the first-order perturbation matrix and the second-order contributions over all possible choices of $\nu$.
And notice that $\lambda =- \omega^2$, and $\delta\lambda = \omega^2 - {\omega_0}^2
\approx 2\omega_0(\delta + \mi \alpha/2)$,
the above equations yield:
\begin{align}
	& (\delta + i\frac{\alpha}{2})\phi = \left(C_\text{1D} + C_\text{rad} + C_\text{2D}\right)\phi \\
	& C_\text{1D}=
	\begin{pmatrix}
		0&\kappa_{2,0}&0&0\\
		\kappa_{-2,0}&0&0&0\\
		0&0&0&\kappa_{0,2}\\
		0&0&\kappa_{0,-2}&0
	\end{pmatrix} \\
	&
	C_\text{2D}=
	\begin{pmatrix}
		{\chi_{y,1,0}^{(1,0)}}&{\chi_{y,1,0}^{(-1,0)}}&
		{\chi_{y,1,0}^{(0,1)}}&{\chi_{y,1,0}^{(0,-1)}}\\
		{\chi_{y,-1,0}^{(1,0)}}&{\chi_{y,-1,0}^{(-1,0)}}&{\chi_{y,-1,0}^{(0,1)}}&{\chi_{y,-1,0}^{(0,-1)}}\\{\chi_{x,0,1}^{(1,0)}}&{\chi_{x,0,1}^{(-1,0)}}&
		{\chi_{x,0,1}^{(0,1)}}&{\chi_{x,0,1}^{(0,-1)}}\\
		{\chi_{x,0,-1}^{(1,0)}}&{\chi_{x,0,-1}^{(-1,0)}}&{\chi_{x,0,-1}^{(0,1)}}&{\chi_{x,0,-1}^{(0,-1)}}
	\end{pmatrix}
	\label{eq:C2D}
	\\
	&
	C_\text{rad}=
	\begin{pmatrix}\zeta_{1,0}^{(1,0)}&\zeta_{1,0}^{(-1,0)}&0&0\\
		\zeta_{-1,0}^{(1,0)}&\zeta_{-1,0}^{(-1,0)}&0&0\\
		0&0&\zeta_{0,1}^{(0,1)}&\zeta_{0,1}^{(0,-1)}\\
		0&0&\zeta_{0,-1}^{(0,1)}&\zeta_{0,-1}^{(0,-1)}
	\end{pmatrix}
	\label{eq:Crad} \\
	& \kappa_{\pm2,0} =
	-\frac{{\omega_0}^{2}}{2\beta_{0}}\xi_{\pm2,0}\int_{PhC}\left|\Theta_{0}\right|^{2}\dif z
	\label{eq:kappa2_0} \\
	& \kappa_{0,\pm 2} =
	-\frac{{\omega_0}^{2}}{2\beta_{0}}\xi_{0,\pm 2}\int_{PhC}\left|\Theta_{0}\right|^{2}\dif z
	\label{eq:kappa0_2} \\
	& \zeta_{p,q}^{(r,s)}=
	-\frac{{\omega_0}^4}{2\beta_0}\iint_{PhC}\xi_{p,q}\xi_{-r,-s}
	G(z,z^{^{\prime}})\Theta_0(z^{^{\prime}})\Theta_0^*(z)dz^{^{\prime}}\dif z
	\label{eq:Crad_addition} \\
	& \chi_{i,p,q}^{(r,s)}=
	-\frac{{\omega_0}^{2}}{2\beta_{0}}\sum_{\sqrt{m^{2}+n^{2}}>1}
	\xi_{p-m,q-n} \varsigma_{i,m,n}^{(r,s)},\quad i=x,y
	\label{eq:C2D_addition}
\end{align}
where $C_\text{1D}$ is the result derived from Eqs.~(\ref{eq:gen_eigen_perturb_raw}-\ref{eq:gen_eigen_1st_order_int}),
$C_\text{2D}$ and $C_\text{rad}$ correspond to the results of
Eqs.~\ref{eq:gen_eigen_2nd_order_mat})-(\ref{eq:gen_eigen_2nd_order_non_expand}
with $\nu \neq (0, 0)$ and $\nu = (0, 0)$ respectively.

For consistency with established literature\cite{RN53,RN54,RN227}, we define the four components of the vector $\phi = (R_x, S_x, R_y, S_y)^T$,
which represent the modes $(1, 0), (-1, 0), (0, 1), (0, -1)$ , respectively.
When finite-size effects are taken into account,the spatial dependence of these four basis vectors must be considered, leading to the finite-size 3D-CWT equation:
\begin{equation}\label{eq:finite_CWT}
	\left(\delta+\mi\frac{\alpha}{2}\right)
	\begin{pmatrix}
		R_x \\
		S_x \\
		R_y \\
		S_y
	\end{pmatrix}
	=
	\left(C_{\text{1D}}+C_{\text{2D}}+C_{\text{rad}}\right)
	\begin{pmatrix}
		R_x \\
		S_x \\
		R_y \\
		S_y
	\end{pmatrix}+\mi
	\begin{pmatrix}
		\partial R_x/\partial x \\
		-\partial S_x/\partial x \\
		\partial R_y/\partial y \\
		-\partial S_y/\partial y
	\end{pmatrix}
\end{equation}

\section{Radiation Loss Model Based on 3D-CWT}
In this section, we analyze the radiation loss based on a Hermiticity analysis of Eq.~(\ref{eq:finite_CWT}).
Apply $\phi^\dagger$ both sides of Eq.~(\ref{eq:finite_CWT}) yields:
\begin{equation}\label{eq:rad_loss1}
	\delta+\mi\frac{\alpha}{2}=
	\phi^\dagger C\phi
	+\mi\iint\left(R_x^*\quad S_x^*\quad R_y^*\quad S_y^*\right)
	\begin{pmatrix}
		\partial R_x/\partial x\\
		-\partial S_x/\partial x\\
		\partial R_y/\partial y\\
		-\partial S_y/\partial y
	\end{pmatrix}\dif x\dif y
\end{equation}
Subtracting the complex conjugate of Eq.~(\ref{eq:rad_loss1}) from itself gives:
\begin{equation}\label{eq:rad_loss2}
	\mi\alpha= \phi^\dagger (C-C^\dagger)\phi
	+\mi\iint\left(
	\frac{\partial\left|R_{x}\right|^{2}}{\partial x}-\frac{\partial\left|S_{x}\right|^{2}}{\partial x}+
	\frac{\partial\left|R_{y}\right|^{2}}{\partial y}-\frac{\partial\left|S_{y}\right|^{2}}{\partial y}
	\right)\dif x\dif y
\end{equation}

It is easy to verify that $C_\text{1D} = C_\text{1D}^\dagger$ and $C_\text{2D}=C_\text{2D}^\dagger$, confirming both are Hermitian matrices
\footnote{This observation has a neat physics interpretation that, for a infinite PCSEL device with lossless
	material, all optical loss comes from the vertical radiation}.
Consequently, $C-C^\dagger = C_\text{rad} - C_\text{rad}^\dagger$, and we obtain:
\begin{equation}\label{eq:rad_loss3}
	\begin{aligned}
		\alpha=
		&2\kappa_{\nu i}
		\iint
		\left((\left|\xi_{-1,0}R_{x}+\xi_{1,0}S_{x}\right|^{2}
		+\left|\xi_{0,-1}R_{y}+\xi_{0,1}S_{y}\right|^{2}))\right)\dif x\dif y\\
		&+\iint
		\left(\frac{\partial\left|R_{x}\right|^{2}}{\partial x}-\frac{\partial\left|S_{x}\right|^{2}}{\partial x}+
		\frac{\partial\left|R_{y}\right|^{2}}{\partial y}-\frac{\partial\left|S_{y}\right|^{2}}{\partial y}\right)\dif x\dif y
	\end{aligned}
\end{equation}
Upon imposing the boundary conditions, Eq.~(\ref{eq:rad_loss3}) simplifies to:
\begin{equation}\label{eq:rad_loss4}
	\begin{aligned}
		\alpha=&
		2\kappa_{\nu,i}\iint(
		\left|\xi_{-1,0}R_{x}+\xi_{1,0}S_{x}\right|^{2}+\left|\xi_{0,-1}R_{y}+\xi_{0,1}S_{y}\right|^{2}
		)\dif x\dif y\\
		&+\int_{0}^{L}\left(\left|R_{ex}(y)\right|^{2}+\left|S_{ex}(y)\right|^{2}\right)\dif y
		+\int_{0}^{L}\left(\left|R_{ey}(x)\right|^{2}+\left|S_{ey}(x)\right|^{2}\right)\dif x
	\end{aligned}
\end{equation}
where $R_{ex}$, $R_{ey}$, $S_{ex}$ and $S_{ey}$ denote the unconstrained boundary values arising from the open boundary condition:
\begin{equation}\label{eq:rad_loss5}
	\begin{aligned}
		&R_{ex}(y) = R_x|_{x=L}\quad S_{ex}(y) = S_x|_{x=0}\\
		&R_{ey}(x) = R_y|_{y=L} \quad S_{ey}(x) = S_y|_{y=0}
	\end{aligned}
\end{equation}

Eq.~(\ref{eq:rad_loss4}) constitutes the final radiation loss model:
the first term on the right-hand side accounts for surface radiation loss,
while the latter two terms account for edge radiation loss, namely:
\begin{align}
	\alpha_{\perp}&=2\kappa_{\nu,i}\iint
	\left(\left|\xi_{-1,0}R_{x}+\xi_{1,0}S_{x}\right|^{2}+\left|\xi_{0,-1}R_{y}+\xi_{0,1}S_{y}\right|^{2}\right)
	\dif x\dif y
	\label{eq:rad_surface}
	\\
	\alpha_{\parallel}&=
	\int_{0}^{L}\left(
	\left|R_{ex}(y)\right|^{2}+\left|S_{ex}(y)\right|^{2}
	\right)\dif y
	+\int_{0}^{L}\left(
	\left|R_{ey}(x)\right|^{2}+\left|S_{ey}(x)\right|^{2}
	\right)\dif x
	\label{eq:rad_edge}
\end{align}

\begin{figure}[htbp]
	\includegraphics[width=1\linewidth]{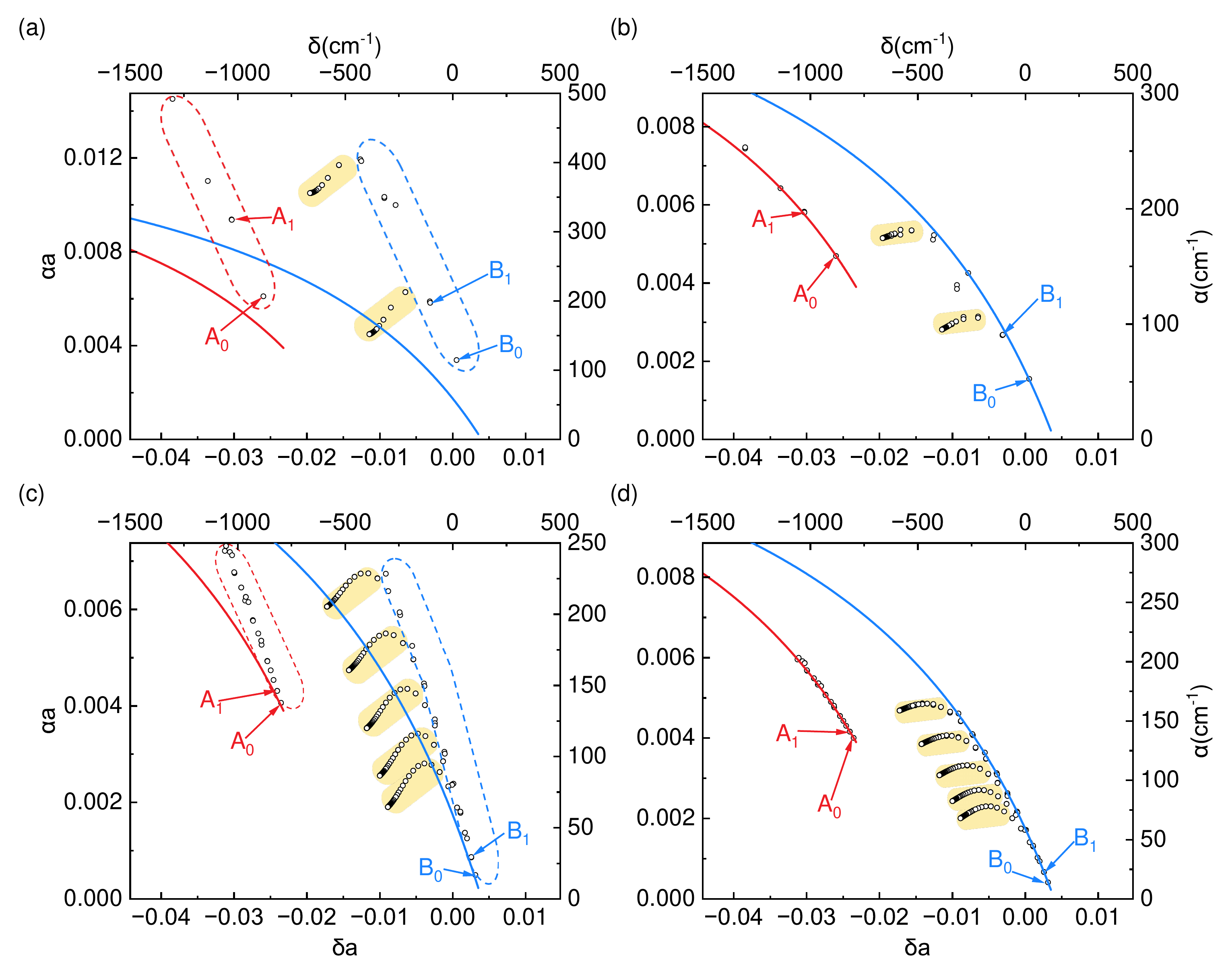}
	\caption{Mode spectrum.
		\textbf{(a)} The mode spectrum of a PCSEL device with an isosceles right-triangular unit cell and size \modify{$L=200a$}.
		Resonance modes $\mathrm{A}$ and $\mathrm{B}$ are marked by red and blue dashed circles, respectively.
		Data points highlighted in yellow are nonphysical artifacts arising from discretizations\cite{RN148}.
		The solid red and blue lines depict the parametric plot of $\delta(k)-\alpha(k)$
		for modes $\mathrm{A}$ and $\mathrm{B}$.
		\textbf{(b)} The mode spectrum after eliminating edge radiation loss for each mode in (a).
		\textbf{(c)} The mode spectrum of a finite-size PCSEL with $L=600a$.
		\textbf{(d)} The mode spectrum after eliminating edge radiation loss for each mode in (c).
	}
	\label{fig:mode_spectrum_L200L600}
\end{figure}

The mode spectra can be naturally obtained by solving Eq.~(\ref{eq:finite_CWT}) using the finite-difference method.
Figs.~\ref{fig:mode_spectrum_L200L600} (a) and (c)
present the mode spectra of PCSELs with isosceles right-triangle unit cells
and device dimensions $L=200a$ and $L=600a$, respectively.
Resonance modes A and B are indicated by red and blue circles, respectively.
As the simulated device size increases, the mode spectra exhibit increasingly dense sampling points.
The red and blue curves are derived from the band structure and can be obtained using Eq.(4) in the main text.
In figs.~\ref{fig:mode_spectrum_L200L600} (a) and (c), the radiative loss of each mode (represented by each point) includes both surface radiation loss($\alpha_\perp$) and edge radiation loss($\alpha_\parallel$).
Subtracting $\alpha_\parallel$ from the total radiative loss of each mode in panels (a) and (c) yields the results shown in panels (b) and (d), respectively.

\section{Edge Radiation Loss from the Perspective of the Poynting Vector}
Eq.~(\ref{eq:rad_edge}) can also be derived from the perspective of the Poynting vector.
For the TE mode, the expression for $H_z$ is given by:
\begin{equation}\label{Hz_expression}
	H_z=\frac{\mi}{\omega}\left(\frac{\partial E_y}{\partial x}-\frac{\partial E_x}{\partial y}\right)
\end{equation}
where $\omega$ is the angular frequency of the corresponding mode and $\mu$ is the magnetic permeability.
Since the four elements of the the vector in Eq.~(\ref{eq:rad_edge}) represents the Fourier basis of
$(\pm 1, 0)$ and $(0, \pm 1)$, the spatial field is:
\begin{align}
	& E_y =
	R_x\Theta_0\me^{-\mi\beta_0x}
	+
	S_x\Theta_0\me^{\mi\beta_0x}
	\label{eq:Ey_expression}\\
	& E_x =
	R_y\Theta_0\me^{-\mi\beta_0y}
	+
	S_y\Theta_0\me^{\mi\beta_0y}
	\label{eq:Ex_expression}
\end{align}

Computing the time-averaged Poynting vector and eliminating the high-frequency terms and small quantities yields:
\begin{align}
	&S_1=
	\frac{1}{2}\real\left(E_x\times H_z^*\right)=
	\frac{1}{2}\frac{\beta_0\Theta_0^2}{\omega}
	\left(|R_y|^2-|S_y|^2\right)\hat{y}
	\label{eq:S1}\\
	&S_2=
	\frac{1}{2}\real\left(E_y\times H_z^*\right)=
	\frac{1}{2}\frac{\beta_0\Theta_0^2}{\omega}
	\left(|R_x|^2-|S_x|^2\right)\hat{x}
	\label{eq:S2}
\end{align}
Eq.~(\ref{eq:S1}) describes energy propagation along the $y$-axis,
and Eq.~(\ref{eq:S2}) describes propagation along the $x$-axis.
Integrating Eq.~(\ref{eq:S1}) over the $x$-$z$ plane and Eq.~(\ref{eq:S2}) over the $y$-$z$ plane,
and applying the boundary conditions, yields:
\begin{align}
	& P_1=
	\frac{\beta_0}{2\omega}\int_0^L
	\left(|R_{ey}(x)|^2+|S_{ey}(x)|^2\right)\dif x
	\label{eq:P1}\\
	& P_2=
	\frac{\beta_0}{2\omega}\int_0^L
	\left(|R_{ex}(y)|^2+|S_{ex}(y)|^2\right)\dif y
	\label{eq:P2}
\end{align}
Eqs.~(\ref{eq:P1}) and (\ref{eq:P2}) represent the power radiated in the $y$- and $x$-directions, respectively.
Their sum has the same functional form as Eqs.~(\ref{eq:rad_edge}),
differing only by a multiplicative factor $\beta_0/2\omega$.

\section{The Wave-packet Spreading Rate of a Gaussian Wave}
In this section, we derive the wave-packet spreading velocity of a Gaussian envelope, assuming a given
dispersion relation $\omega(k)$. This derivation is independent of the detail on how the dispersion
$\omega(k)$ is formed and therefore does not depend on the specific structure of the photonic system
or what is discussed above.
\modify{For simplicity, we consider a one-dimensional Gaussian envelope in this derivation (the scaling conclusions are unchanged for a two-dimensional geometry).}
Assuming a Gaussian-shaped intensity profile in k-space is a good approximation of the envelope function
of the eigenmode in a PCSEL device:
\begin{equation}\label{eq:Gaussian_k}
	I( k)=
	\frac{1}{\sqrt{2\pi}\sigma_{k}}
	\exp\left(-\frac{| k|^2}{2\sigma_k^2}\right),
	\quad\sigma_k=\pi/(2L)
\end{equation}
where $\sigma_k$ denotes the width of the Gaussian distribution in $k$-space,
and $L$ represents the device dimension in the simulation.
From Eq.~(\ref{eq:Gaussian_k}), \modify{the amplitude} of the envelope is given by:
\begin{equation}\label{eq:Gaussian_k_A}
	A(k)=
	\left(\frac{1}{2\pi\sigma_k^2}\right)^{\frac{1}{4}}\exp\left(-\frac{\left|k\right|^2}{4\sigma_k^2}\right)
\end{equation}
Approximate $\omega(k)$ up the 2nd order in Taylor expansion, and for simplicity, assuming it is isotropic:
\begin{equation}
	\omega(k)=\overline{\omega}+\frac{1}{2}\omega^{^{\prime\prime}}k^2
\end{equation}
The inverse Fourier transform of Eq.~(\ref{eq:Gaussian_k_A}) evaluated at time $t$ yields:
\begin{equation}\label{eq:Gaussian_rt}
	E(r,t)=
	e^{i\overline{\omega}t}\left(\frac{1}{2\pi\sigma_k^2}\right)^{\frac{1}{4}}\frac{\sigma_k}{\sqrt{\pi}}\frac{1}{\sqrt{1-2i\sigma_k^2\omega^{\prime\prime}t}}e^{-\frac{\sigma_k^2x^2\left(1+i2\sigma_k^2\omega^{\prime\prime}t\right)}{1+(2\sigma_k^2\omega^{\prime\prime}t)^2}}
\end{equation}
The width of the field in Eq.~(\ref{eq:Gaussian_rt}) is:
\begin{equation}\label{eq:sigma_rt}
	\sigma_r^2(t)=
	\frac{1+(2\sigma_k^2\omega''t)^2}{4\sigma_k^2}
\end{equation}
whose time derivative gives:
\begin{equation}\label{eq:deff_sigma_rt}
	\frac{\dif\sigma_r(t)}{\dif t} =
	\frac{{\sigma_k}^2\omega''^2t}{\sigma_r}
\end{equation}
For the fundamental mode, $\sigma_k\sim 1/L$ and the spatial width $\sigma_r\sim L$, therefore:
\begin{equation}\label{eq:deff_sim_sigma_rt}
	\frac{\dif\sigma_r\left(t\right)}{\dif t}\sim
	\frac{\omega''^2t}{L^3}\propto\frac{\omega''^2}{L^3}
	\propto \frac{\delta''^2}{L^3}
\end{equation}
where $\delta^{\prime\prime}=\omega^{\prime\prime}n_{eff}/c$.
Eq.~(\ref{eq:deff_sim_sigma_rt}) indicates that the wave-packet spreading velocity is proportional to $L^{-3}$.

\modify{
The derivation above is not a complete quantitative theory.
Our inspiration partly comes from the work of Persson et al\cite{RN451}.
The Gaussian wave-packet is used solely as a toy model to illustrate the parametric dependence of the lateral leakage on
$L$.
We use $\dif\sigma_r/\dif t$ as a measure because our numerical results (Fig.~8(b)) show that the edge radiation loss scales with ($\omega''^2$), and the spreading rate in Eq.~(\ref{eq:deff_sim_sigma_rt}) intrinsically contains $\omega''^2$.
We expect that other reasonable envelope shapes (e.g., rectangular, sinc, or the actual eigenmode obtained from Eq.(2)) exhibit the same $L^{-3}$ scaling, because the leading spreading dynamics are governed by the band curvature $\omega''$ and the scaling with $L$ follows from $\sigma_k\sim 1/L$ and $\sigma_r\sim L$.
For a physically realistic wave-packet, the first-order (linear) spreading velocity at $t=0$ is zero; only second-order (quadratic) spreading exists.
However, the Gaussian wave-packet is not the true eigenmode of the device.
The actual eigenmode, when spreading towards the boundary, should behave more like a Gaussian at small $t > 0$  rather than at $t = 0$.
At any infinitesimally small positive time, the spreading velocity is non-zero.
The vanishing of the instantaneous spreading rate at $t=0$(Eq.~(\ref{eq:deff_sigma_rt})) is a consequence of the quadratic dispersion and does not affect the final $L^{-3}$ scaling, which is derived from the combination of parameters
$\sigma_k$ and $\sigma_r$.
The ''$t \to 0$'' notation in the main text (Eqs.~(15) - (16)) is used to indicate the parametric scaling rather than a literal zero-time limit.
We emphasize that the Gaussian model is phenomenological;
its validity rests on consistency with numerical simulations (Fig.~8(b)) rather than on a rigorous first-principles derivation.
A full theoretical treatment of the lateral-loss scaling is left for future work.
The numerical results in Fig.~8(b) support the scaling argument despite the toy-model nature of the Gaussian derivation.
}
\section{Finite-size deviation from asymptotic scaling laws}
\modify{ The asymptotic scaling laws $\alpha_\perp-\alpha_{\infty} \sim O(L^{-2})$ and $\alpha_{\parallel} \sim O(L^{-3})$ are strictly valid only in the limit $L \to \infty$.
For finite device sizes, two related deviations are observed.}

\modify{ First, the onset size at which the asymptotic scaling begins to hold depends on the hole geometry: circular holes reach the asymptotic regime already at $L \sim 50\ \mu\text{m}$, isosceles right-triangular (RIT) holes at $L \sim 100\ \mu\text{m}$, and double-lattice structures only at $L \sim 300\ \mu\text{m}$.
Second, for small $L$ the effective exponent extracted from finite-CWT simulations is smaller than the asymptotic value (e.g., Fig.~5 in the main text).}

\modify{Both phenomena originate from the nonparabolicity of the photonic bands away from the $\Gamma$ point.
When $L$ decreases, the mode's $k$-space extent $\Delta k \sim \pi/L$ broadens, sampling regions where the dispersion is no longer quadratic.
The stronger the nonparabolicity, the larger the $L$ required to stay near the parabolic region.}

\modify{To quantify this, we fitted the real and imaginary parts of the band $\omega(k)$ for the $A$ mode up to fourth order in $k$.
The coefficients are listed in Tables.~\ref{tab:real_part} and \ref{tab:real_part}.
The double-lattice structure exhibits an extremely large fourth-order real coefficient ($4.657 \times 10^{6}$), indicating very strong nonparabolicity, which explains its late onset.
Circular holes have the smallest fourth-order coefficient, and RIT holes lie in between.}

\modify{At small $L$, the quartic term becomes non-negligible.
For all three structures, the quadratic and quartic coefficients have opposite signs: for the real part, second-order negative and fourth-order positive; for the imaginary part, second-order positive and fourth-order negative.
Because of this sign opposition, the quartic term reduces the magnitude of $\omega(k)$ relative to the purely quadratic (parabolic) approximation.
Consequently, the computed losses for small devices lie below the asymptotic power-law curves, and the effective exponent is smaller.
As $L$ increases, the sampled $k$ shrinks, the quartic term becomes negligible, and the data approach the asymptotic power laws.}

\modify{Thus, the different validity ranges and the reduced exponent at small $L$ share a common physical origin: band nonparabolicity and the sign opposition of the expansion coefficients.}

\begin{table}[htbp]
	\centering
	\caption{Fitting coefficients of the real part of the dispersion relation}
	\label{tab:real_part}
	\begin{tabular}{lcc}
		\toprule
		Shape & Mode A 2nd-order Fitting & Mode A 4th-order Fitting \\
		& Coefficient & Coefficient \\
		\midrule
		Circular & -11.77 & $3.058 \times 10^{3}$ \\
		Isosceles Right Triangle & -18.71 & $1.433 \times 10^{4}$ \\
		Double-lattice & -142.9 & $4.657 \times 10^{6}$ \\
		\bottomrule
	\end{tabular}
\end{table}

\begin{table}[htbp]
	\centering
	\caption{Fitting coefficients of the imaginary part of the dispersion relation}
	\label{tab:imaginary_part}
	\begin{tabular}{lcc}
		\toprule
		Shape & Mode A 2nd-order Fitting & Mode A 4th-order Fitting \\
		& Coefficient & Coefficient \\
		\midrule
		Circular & $4.392 \times 10^{5}$ & $-4.269 \times 10^{8}$ \\
		Isosceles Right Triangle & $6.745 \times 10^{5}$ & $-1.646 \times 10^{9}$ \\
		Double-lattice & $5.590 \times 10^{6}$ & $-4.577 \times 10^{11}$ \\
		\bottomrule
	\end{tabular}
\end{table}

\bibliography{pcsel_sup}
